\documentclass[a4paper,11pt]{article}

\usepackage{jinstpub} 

\usepackage{lineno}

\title{\boldmath X-ray Performance of a Small Pixel Size sCMOS Sensor and the Effect of Depletion Depth}

\author[a,b]{Yu Hsiao,}
\author[a,b,1]{Zhixing Ling,\note{Corresponding author}}
\author[a,b]{Chen Zhang,}
\author[a]{Wenxin Wang,}
\author[d]{Quan Zhou,}
\author[d]{Xinyang Wang,}
\author[a,b,c]{Shuang-Nan Zhang}
\author[a,b]{and Weimin Yuan}

\affiliation[a]{National Astronomical Observations, Chinese Academy of Sciences,\\20A Datun Road, Beijing, 100101, China}
\affiliation[b]{School of Astronomy and Space Science, University of Chinese Academy of Science,\\19A Yuquan Road, Beijing, 100049, China}
\affiliation[c]{Institute of High Energy Physics, Chinese Academy of Sciences,\\19B Yuquan Road, Shĳingshan District, Beijing, 100049, China}
\affiliation[d]{Gpixel Inc.,\\No.588 Yingkou Road, Economical and Technological Development Zone, Changchun, 13033, China}

\emailAdd{lingzhixing@nao.cas.cn}

\abstract{In recent years, scientific Complementary Metal Oxide Semiconductor (sCMOS) devices have been increasingly applied in X-ray detection, thanks to their attributes such as high frame rate, low dark current, high radiation tolerance and low readout noise. We tested the basic performance of a backside-illuminated (BSI) sCMOS sensor, which has a small pixel size of $\rm{6.5\ \mu m\times6.5\ \mu m}$. At a temperature of -20$^\circ \rm{C}$, The readout noise is 1.6 $\rm{e^-}$, the dark current is 0.5 $\rm{e^-/pixel/s}$, and the energy resolution reaches 204.6 eV for single-pixel events. The effect of depletion depth on the sensor's performance was also examined, using three versions of the sensors with different deletion depths. We found that the sensor with a deeper depletion region can achieve a better energy resolution for events of all types of pixel splitting patterns, and has a higher efficiency in collecting photoelectrons produced by X-ray photons. We further study the effect of depletion depth on charge diffusion with a center-of-gravity (CG) model. Based on this work, a highly depleted sCMOS is recommended for applications of soft X-ray spectroscopy.}


\begin{document}
\maketitle
\flushbottom

\section{Introduction}
\label{sec:intro}
In recent years, scientific complementary mental-oxide-semiconductor (sCMOS) sensors have increasingly been used for X-ray applications in a wide range of fields. Compared to CCDs, the conventional imaging sensors, sCMOS sensors feature higher frame rate and lower readout noise, making them an interesting alternative to X-ray detection. Scientific CMOS cameras have been used in hard X-ray tomography~\cite{i} and soft X-ray beamlines~\cite{e}. Recently sCMOS sensors start to find applications in X-ray astronomy, thanks to their good performance in X-ray imaging and spectroscopy. For a few astronomical missions under development or mission proposals, e.g. FOXISI3~\cite{f} for solar X-ray observation, Einstein Probe (EP)~\cite{m}\cite{n} and THESEUS~\cite{g} for detecting soft X-ray transients, sCMOS detectors are adopted. Hybrid CMOS detectors with high throughput and deep depletion depth are also developed for future X-ray missions such as Lynx~\cite{k}. 

We have been working on sCMOS for X-ray application in astronomy since 2015, and for optimizing and customizing CMOS X-ray detectors of high quality~\cite{j}. We also found and studied the crosstalk phenomenon of X-ray events in an sCMOS sensor with a pixel size of $\rm{11\ \mu m\times11\ \mu m}$~\cite{h}. In this study, we investigated the X-ray performance of a small pixel-size backside-illuminated (BSI) sCMOS sensor GSENSE2020BSI (G2020 for short)\footnote{\url{https://www.gpixel.com/products/area-scan-en/gsense/gsense2020bsi/}}. 
We test its basic properties and capability in X-ray spectroscopy. Ideally, the signal can be well restored if the incident charge is collected within the depleted region of the sensor. However, it is still a challenge for today's sCMOS technology to manufacture sensors with sufficiently thick depletion depth. It is therefore of importance to evaluate the effect of depletion depth on the X-ray performance of CMOS. To this end we produced three versions of G2020 with the same epitaxial layer of thickness 10 $\rm{\mu m}$ but different depletion depths, or in other words, different doping densities. One of these sensors is fully depleted (designated by H) and the other two are partially depleted (denoted by M and L for the medium and minimum level of depletion, respectively). We carried out a series of tests to study how the depletion depth affects the sensors' X-ray performance. We conclude that a sensor with a deeper depletion region has better performance in collecting charge and spectrum analysis. Based on the results, we come up with a center-of-gravity model~\cite{a} to estimate the depletion depth, which is expected to be applied in calculating the depletion depth of sCMOS sensors in general.

The basic characteristics of the sCMOS sensor G2020 are described in Section~\ref{sec:basics}. The X-ray performance of the sensors and the depletion depth estimation are discussed in Section~\ref{sec:X-ray}, followed by conclusions in Section~\ref{sec:conclusions}. 

\section{Characteristics of GSENSE2020BSI}
\label{sec:basics}
\begin{figure}[htbp]
\centering
\includegraphics[width=.7\textwidth]{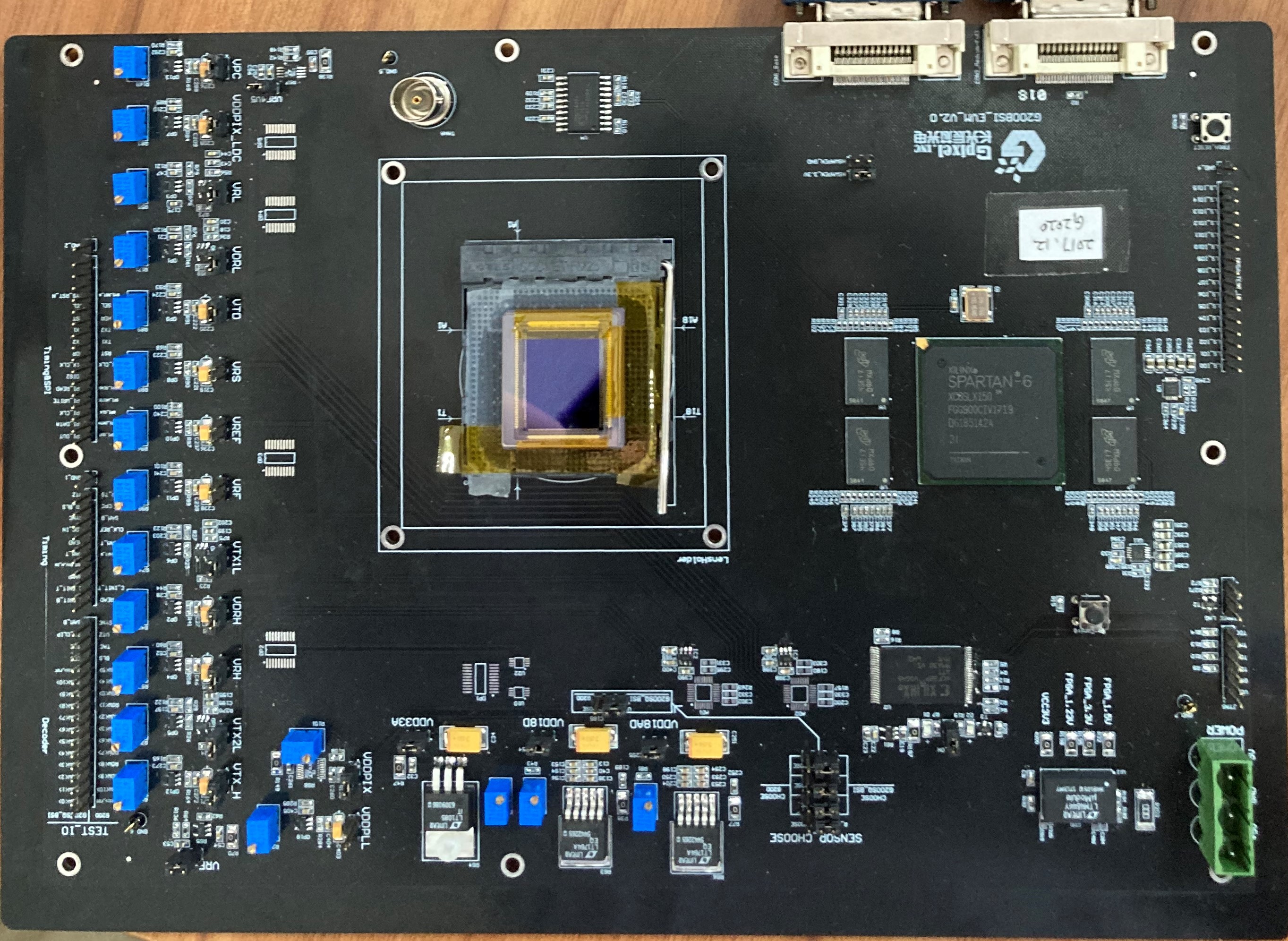}
\caption{\label{fig:eva_board}The evaluation board used to test GSENSE2020BSI.}
\end{figure}

The standard backside illuminated sCMOS image sensor used in this study, namely G2020, has $2048\times2048$ active pixels, each of $\rm{6.5\ \mu m \times 6.5\ \mu m}$ in size. G2020 has a six-transistor pixel architecture. The basic properties of G2020 are listed in Table~\ref{tab:i}. Note that the properties described below refer to the standard fully depleted sensor H unless specified otherwise. An official evaluation board is used in our experiment. The picture of the evaluation board and the sensor is shown in Figure~\ref{fig:eva_board}. 

The other two sensors were produced in the same fabrication method except for a different doping density during the epi growing. The estimated depth of the depletion layer of the type H sensor is around 10 $\rm{\mu m}$ according to the data sheet of the fab, which is the thickness of the epitaxial layer. This means that the epi is fully depleted. The estimation of the depletion depth for the two partially depleted sensors will be discussed in Section~\ref{sec:X-ray}. Note that the three sensors have equal thickness of their epitaxial layers, which guarantees the same quantum efficiency (QE) for X-ray detection.

\begin{table}[htbp]
\centering
\caption{\label{tab:i} Basic properties of GSENSE2020BSI.}
\smallskip
\begin{tabular}{lrc}
\hline
item&value\\
\hline
Pixel size & $\rm{6.5\ \mu m \times 6.5\ \mu m}$\\
Number of active pixels & $2048\times2048$\\
Photosensitive area & $13.3\ {\rm mm }\times 13.3\ {\rm mm }$\\
Peak QE (at 550nm)& 95$\%$\\
Maximum frame rate & 43 fps @ Rolling HDR mode 12 bit\\
 & 74 fps @ Rolling HDR mode 11 bit\\
Full well capacity & $\rm{55\ ke^-}$\\
Dynamic Range & $90\ \rm{dB}$ @Rolling HDR\\
Dark current & $\rm{80\ e^-/pixel/sec}$ @ 35$^{\circ}\rm{C}$\\
 & $\rm{0.16\ e^-/pixel/sec}$ @ -20$^{\circ}\rm{C}$\\
\hline
\end{tabular}
\end{table}

\subsection{Fixed pattern noise (FPN) and readout noise}
\begin{figure}[htbp]
\centering 
\includegraphics[width=.45\textwidth]{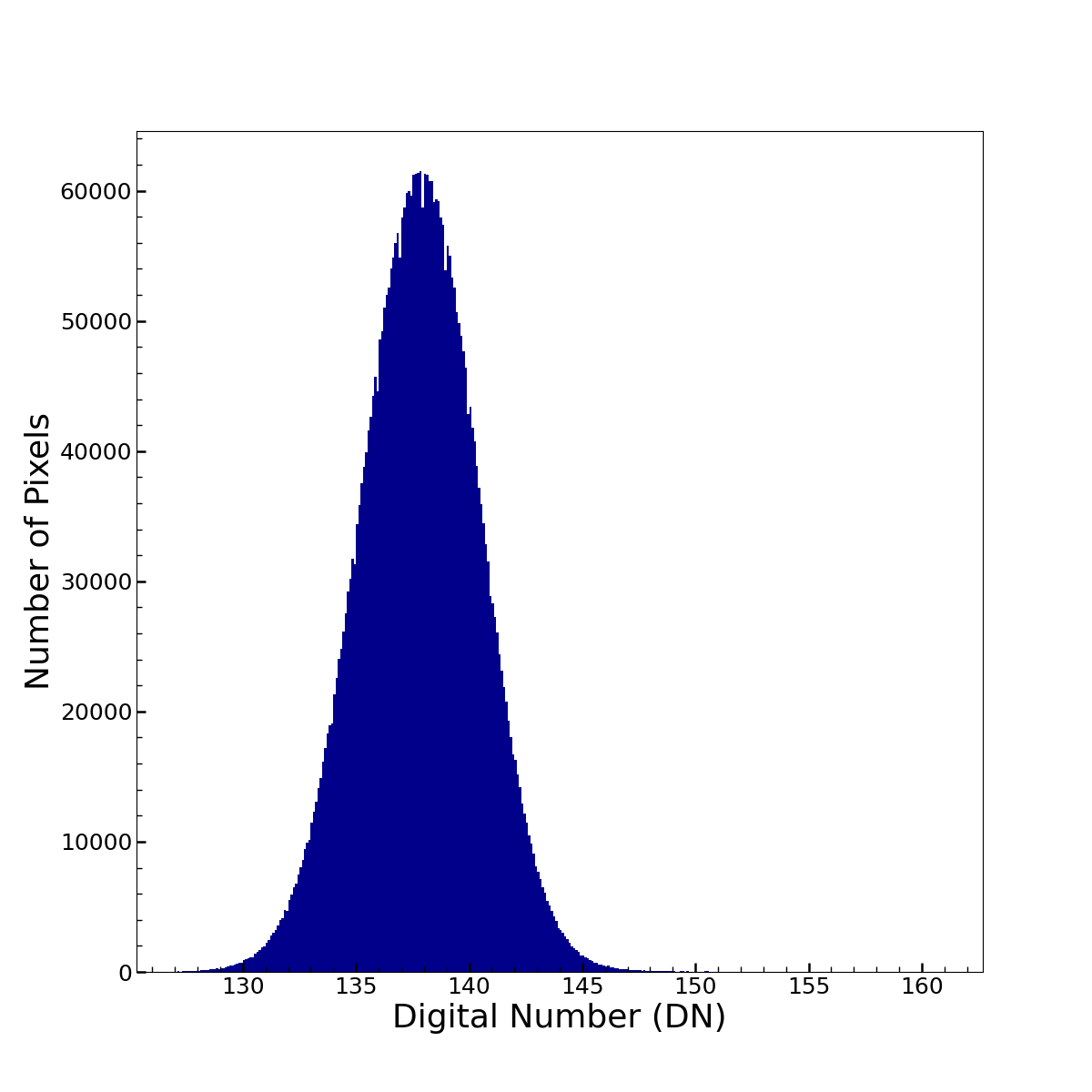}
\quad
\includegraphics[width=.45\textwidth]{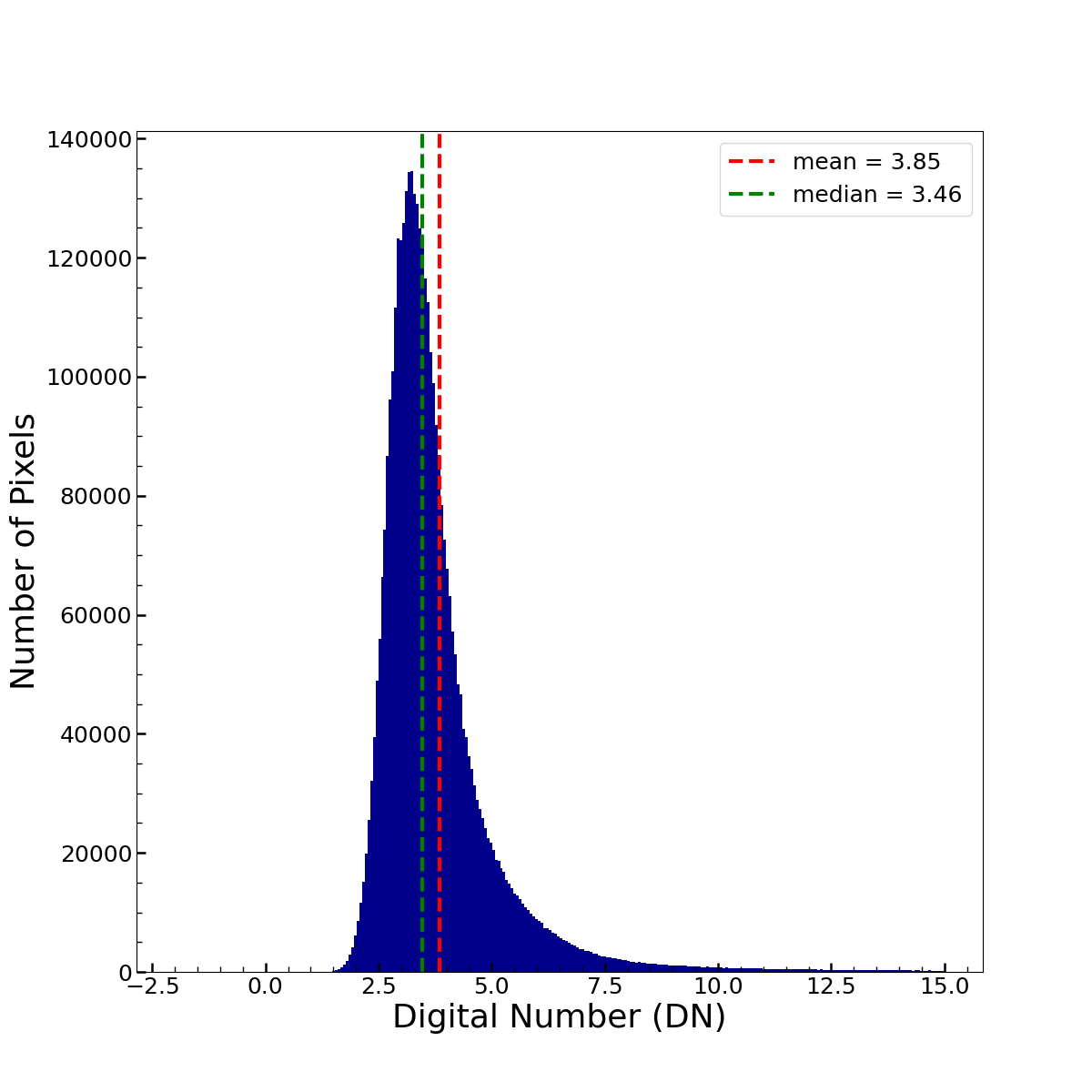}
\caption{\label{fig:1} Distribution of the bias (left) and the noise (right) of type H sensor at -20$^\circ \rm{C}$. The exposure time is set to 5.64 $\rm{\mu s}$.}
\end{figure}

\begin{figure}[htbp]
\centering
\includegraphics[width=.45\textwidth]{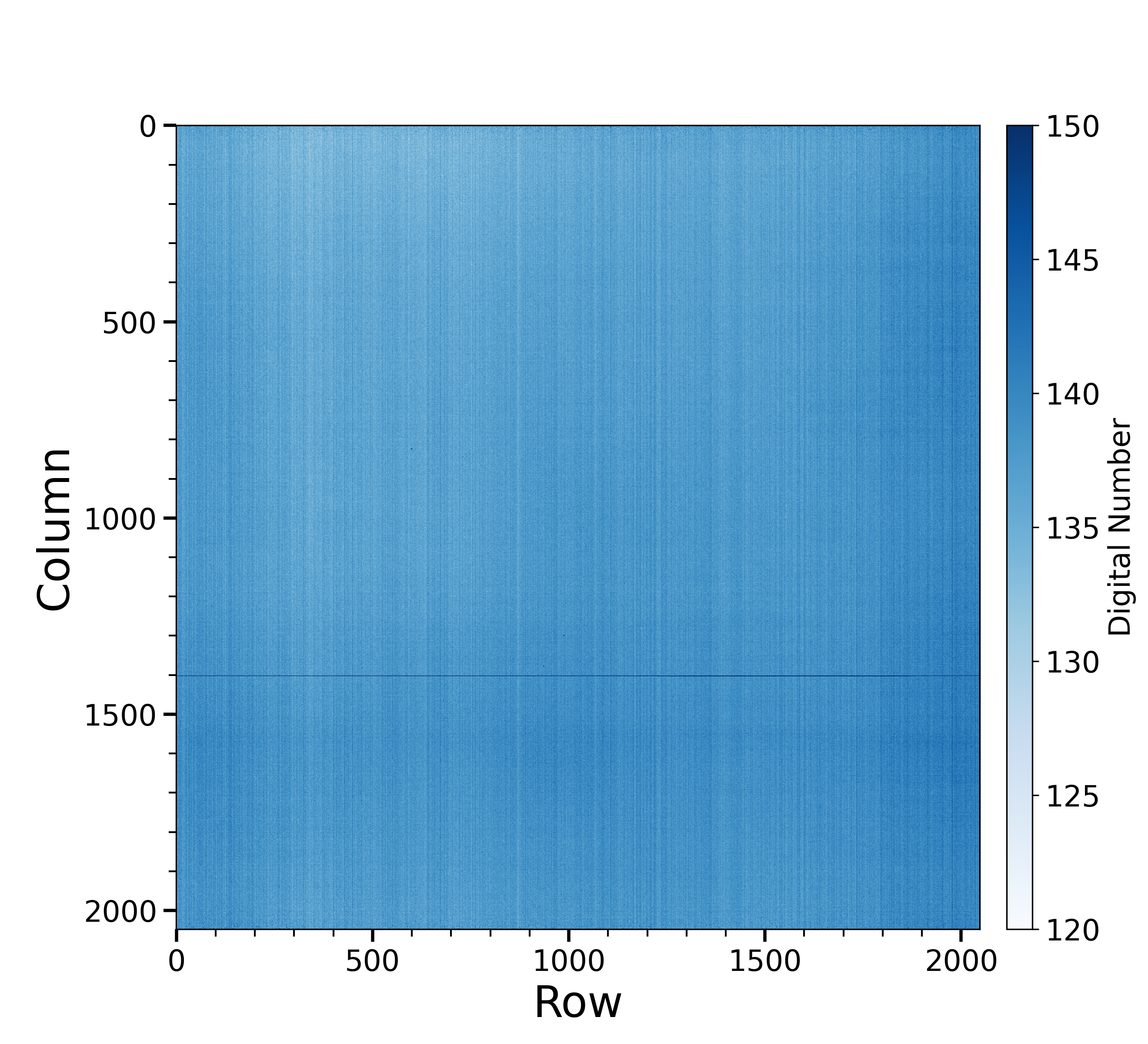}
\quad
\includegraphics[width=.45\textwidth]{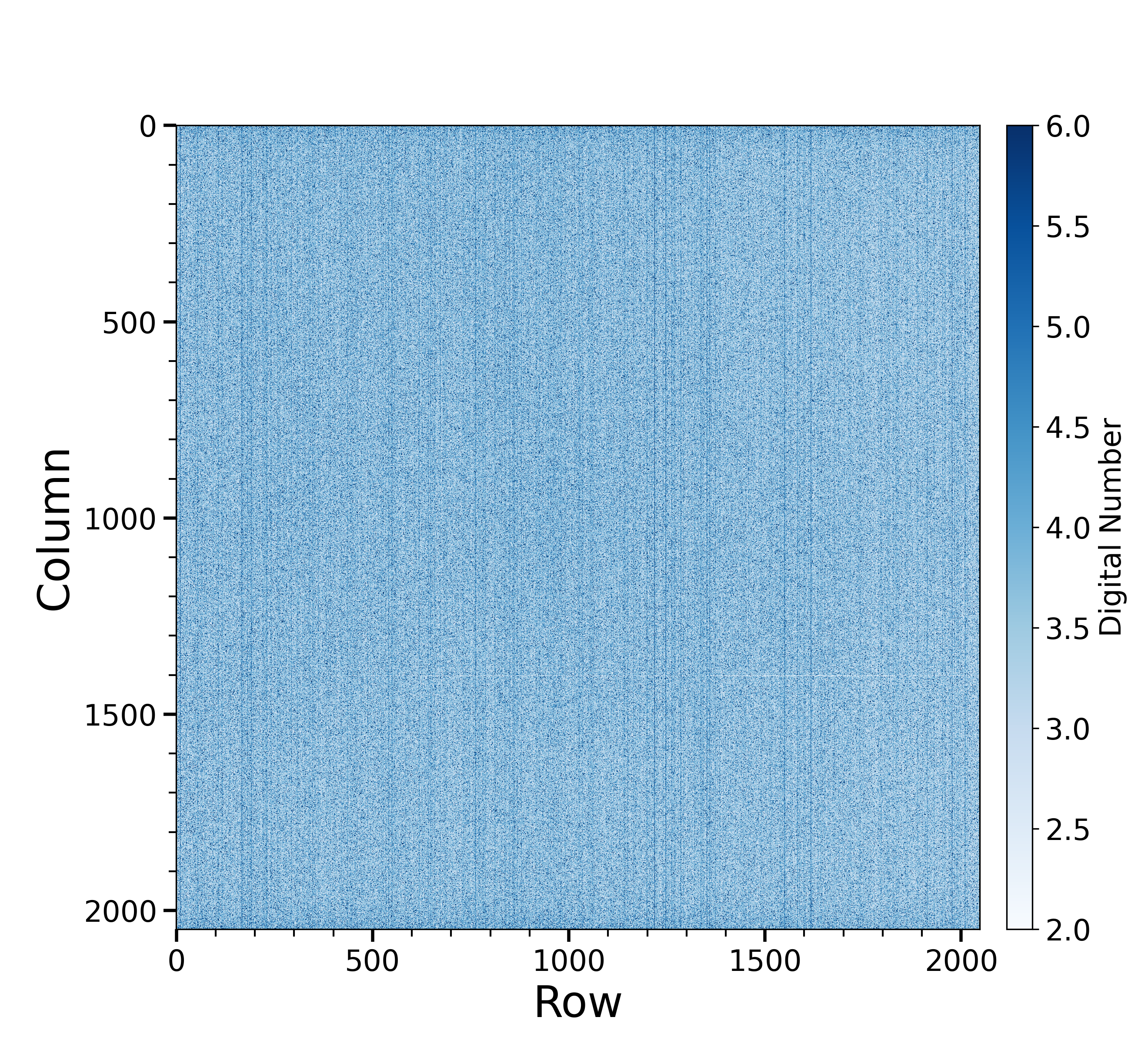}
\caption{\label{fig:2} Maps of the bias (left) and the noise (right) of type H sensor at -20$^\circ \rm{C}$. The exposure time is set to 5.64 $\rm{\mu s}$.}
\end{figure}

\begin{figure}[htbp]
\centering 
\includegraphics[width=.7\textwidth]{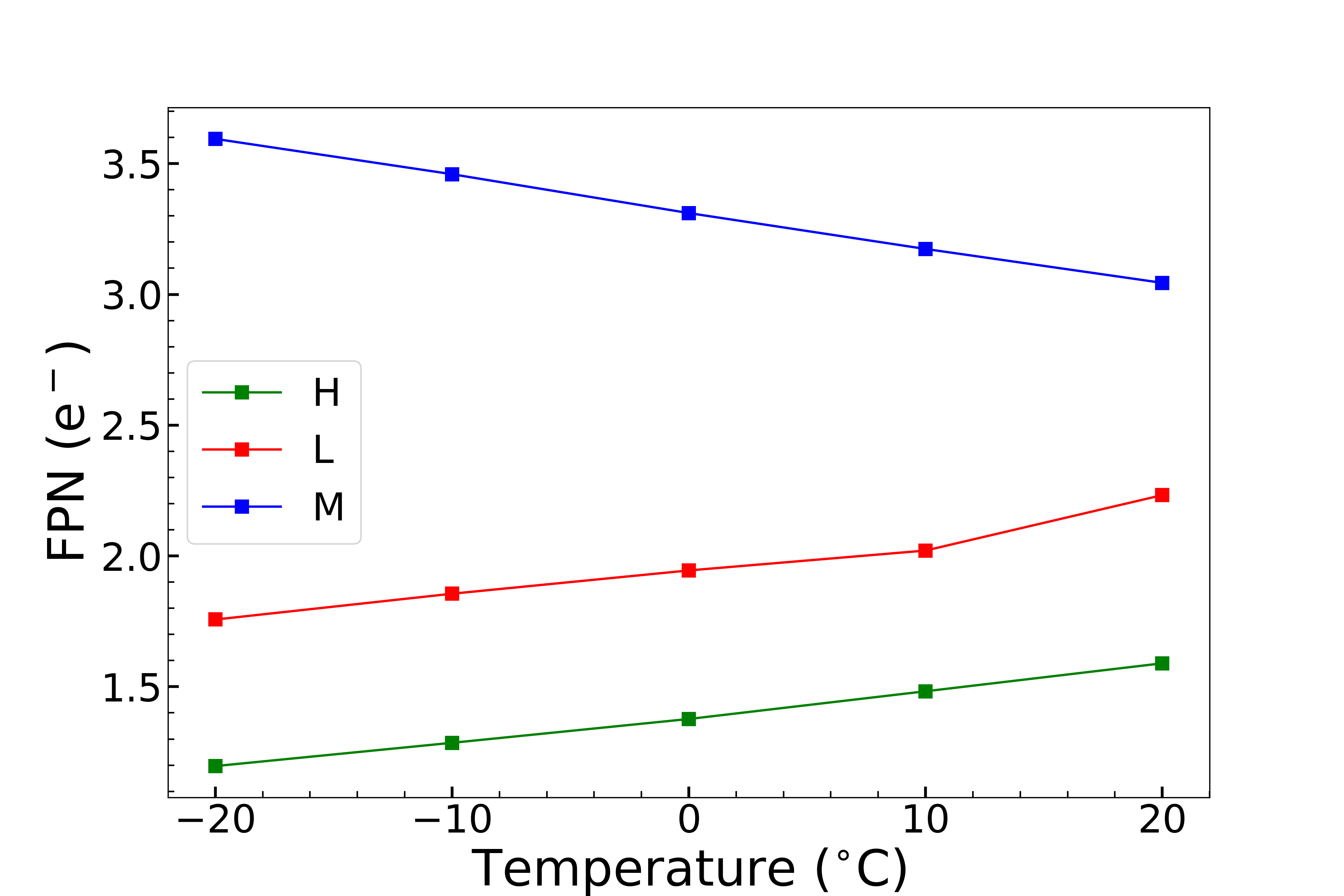}
\caption{\label{fig:FPN} Dependence of FPN on temperature for the three sensors with different depletion depths. The conversion gain is 1.67 eV/DN, 1.50 eV/DN and 1.66 eV/DN for the type H, type M and type L sensor, respectively}
\end{figure}

\begin{figure}[htbp]
\centering
\includegraphics[width=.7\textwidth]{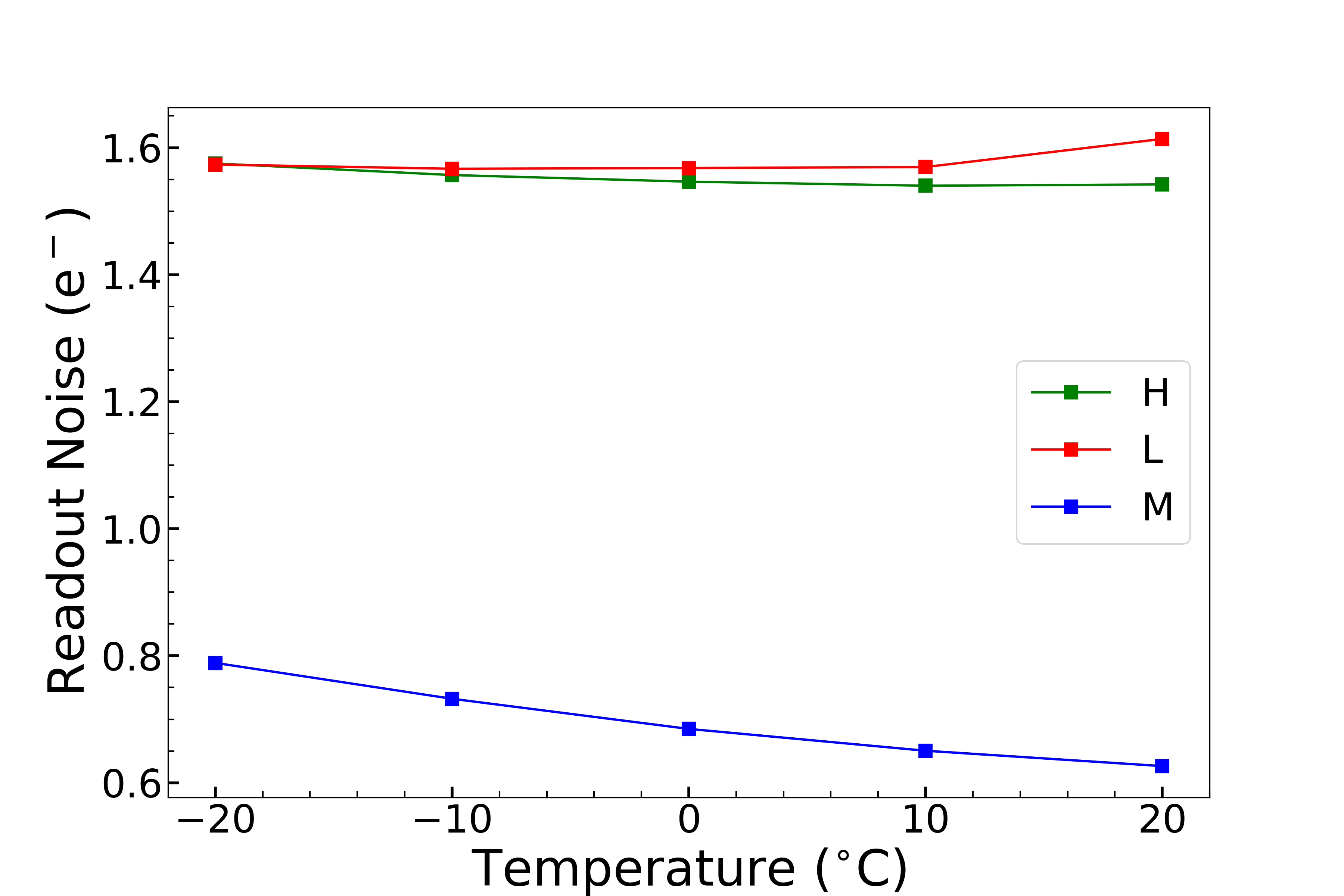}
\caption{\label{fig:RN} Dependence of readout noise on temperature of the three sensors with different depletion depth.}
\end{figure}

We measured the fixed pattern noise (FPN), readout noise and dark current of the three sensors in a dark chamber with the shortest possible integration time 5.64 $\rm{\mu s}$. The temperature was set between -20$^\circ \rm{C}$ and 20$^\circ \rm{C}$ with an interval of 10$^\circ \rm{C}$. The temperature mentioned below all refers to the environment temperature in the chamber. Then we calculated the mean and standard deviation of each pixel using 50 dark frames acquired at each of the working temperatures. Taking the measurement at -20$^\circ \rm{C}$ of the type H sensor as an example, the distribution of the mean value of all the pixels is shown in the left panel of Figure~\ref{fig:1}. The standard deviation of this distribution is 3.84 DN, which indicates the FPN. The right panel of Figure~\ref{fig:1} shows the distribution of the standard deviations over the 50 images of all the pixels. Considering the significant tail in this distribution, the median ($\sim$3.46 DN) rather than the mean is used to represent the readout noise. Figure~\ref{fig:2} shows the maps of the bias and noise corresponding to the distribution above. The noise levels at different temperatures for each sensor are shown in Figure~\ref{fig:FPN} and Figure~\ref{fig:RN}. By using the conversion gain at -20$^\circ \rm{C}$, which is acquired in Section~\ref{sec:X-ray}, the readout noises of the sensor H and L are calculated to be approximately 1.6 $\rm{e^-}$, which are consistent with the official data. The FPN of the type H and type L sensor increases slightly as the temperature rises. The unusual noise level appeared in the sensor M is probably due to some defect during fabrication.

\subsection{Dark current}
\begin{figure}[htbp]
\centering
\includegraphics[width=.7\textwidth]{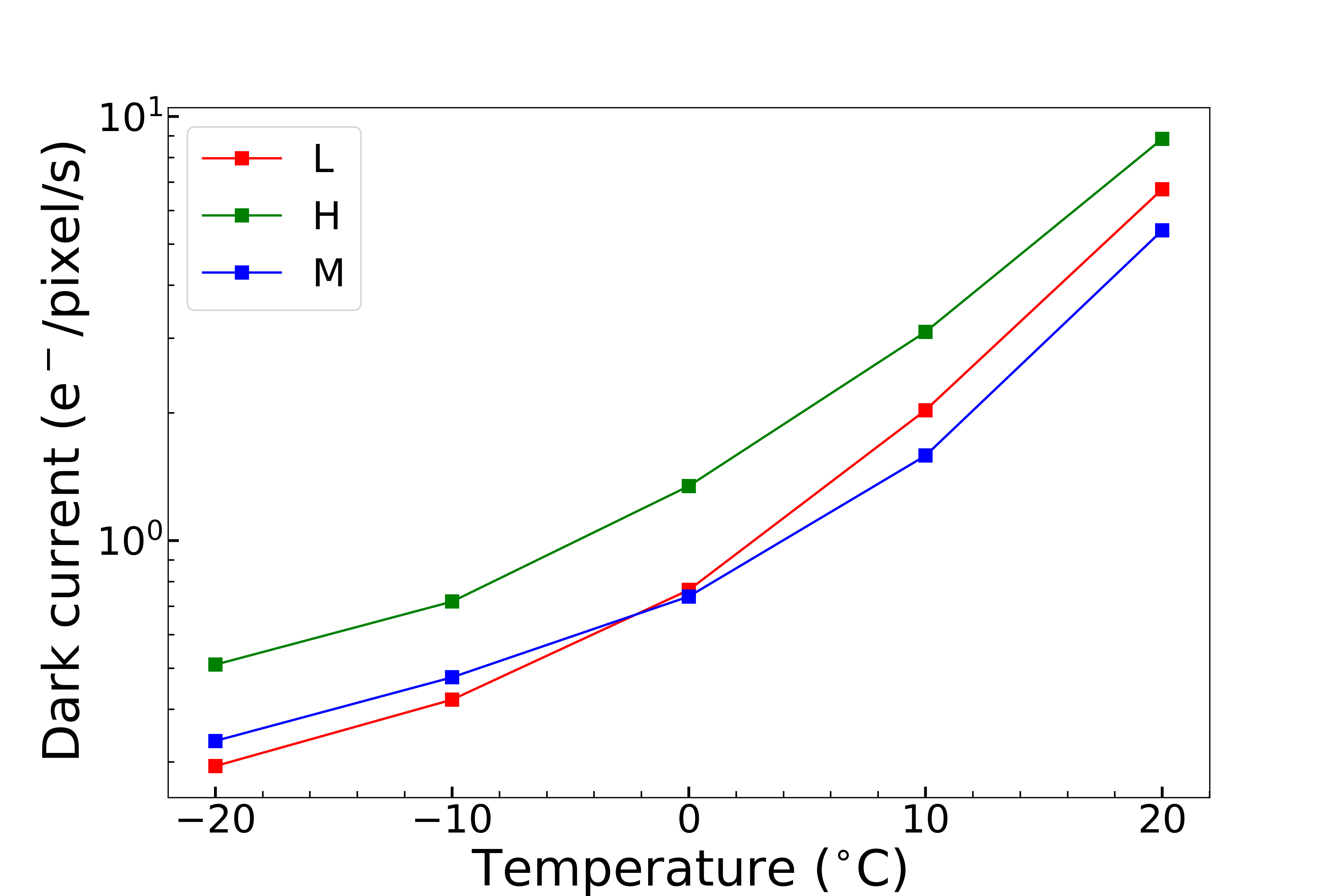}
\caption{\label{fig:DC} Dependence of the dark current on temperature of the three sensors with different depletion depths.}
\end{figure}

Dark current is caused by random thermal excitation of electrons in the depletion area. To measure the dark current level for a given temperature, we took several dark frames with integration time ranging from the shortest exposure time (5.64 ms) to 100 s. For each integration time, the mean value of 50 images taken is determined for each pixel and the median of all these pixels is used to represent the dark charge. We then apply a linear fit to model the relation between the dark charges and integration time. The fitted slope of the linear relation gives the dark current at that temperature. The dark currents for the three sensors at different temperature (from -20$^\circ \rm{C}$ to 20 $^\circ \rm{C}$) are displayed in Figure~\ref{fig:DC}. All the sensors show significant decrease in dark current as the temperature goes down, reaching 0.5, 0.2 and 0.1 $\rm{e^-/pixel/s}$ for sensor H, M and L, respectively, at -20$^\circ \rm{C}$. The similar value and tendency mean that the different depletion depths make little difference to a sensor's dark current.

\section{X-ray performance of GSENSE2020BSI}
\label{sec:X-ray}
\subsection{Event distribution and X-ray spectrum}
\label{sec:X-ray_1}
An $^{55}$Fe X-ray source is used to test the energy response of GSENSE2020BSI. The results of the three sensors were obtained with temperature controlled at -20$^\circ \rm{C}$. The position of the source and the sensor on the evaluation board were fixed during the whole test.

\begin{figure}[htbp]
\centering
\includegraphics[width=.8\textwidth]{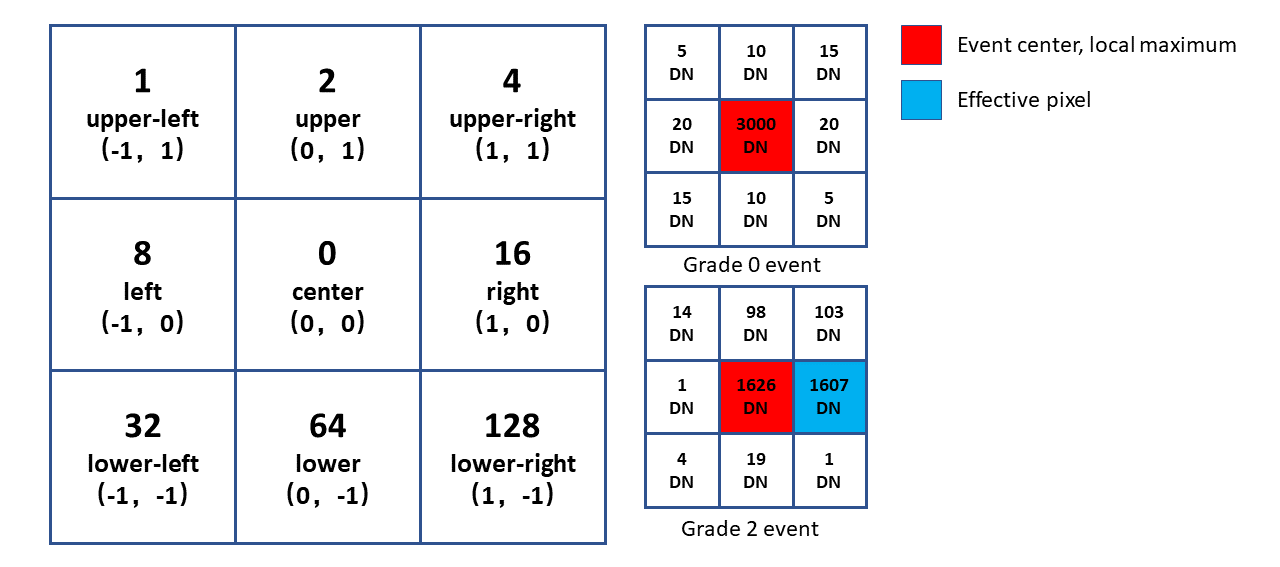}
\caption{\label{fig:grade} Grade definition(Adapted from Ling et al. 2021~\cite{h}).}
\end{figure}

We used the data processing method based on Ling et al. 2021~\cite{h}. The bias generated by the first 50 images is subtracted from all the images. Then we scanned all pixels to select X-ray events with the threshold set at $T_{\rm event} = 100$ DN. If the charge in a pixel is found to exceed this threshold and is the local maximum over its adjacent $3\times3$ pixel region, the digital number in this $3\times3$ area will be recorded as an event. The $3\times3$ window is large enough to collect all the charge produced by a single photon for the fully depleted sensor H. The split pattern can be different from one event to another. Therefore for the $3\times3$ data, we follow the grade scheme for the ACIS (Advanced CCD Imaging Spectrometer) of the Chandra X-ray telescope \cite{l}, as shown in Figure~\ref{fig:grade}. A split threshold $T_{\rm split} = 50$ DN is defined to assign a Grade value from 0 to 255 to each event.

\begin{figure}[htbp]
\centering
\includegraphics[width=.45\textwidth]{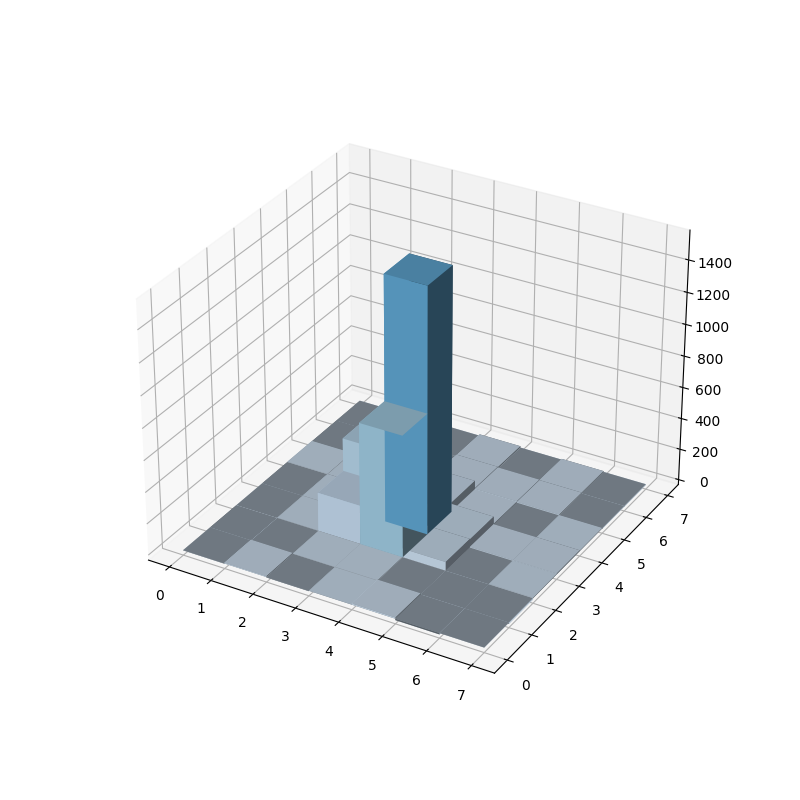}
\quad
\includegraphics[width=.45\textwidth]{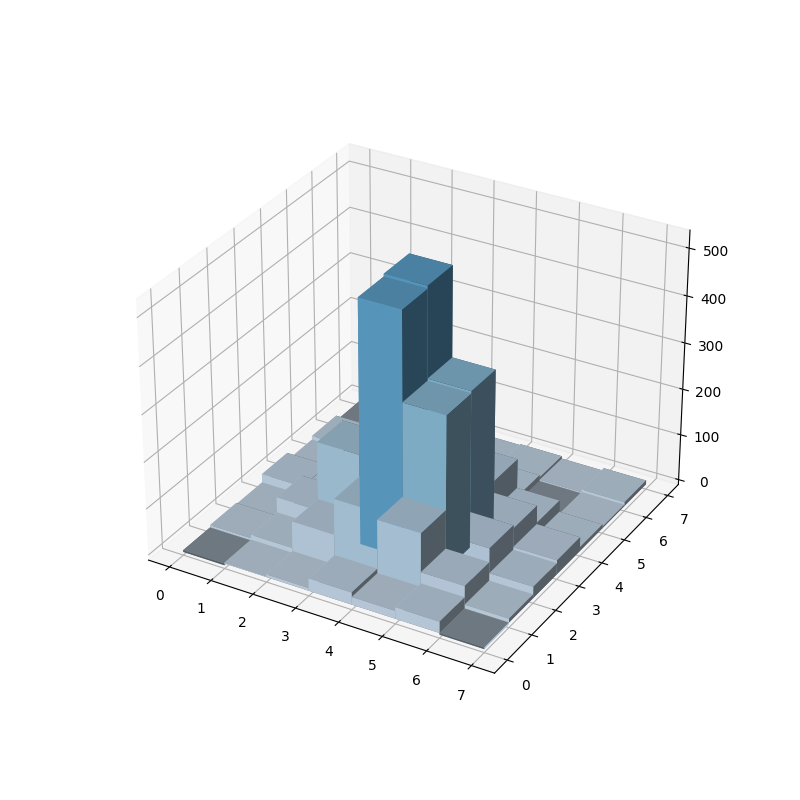}
\caption{\label{fig:Bar3d} Left: A typical X-ray event recorded by sensor H. electrons are collected within a $3\times3$ region; Right: A typical X-ray event recorded by sensor L with electrons split into a $7\times7$ area.}
\end{figure}

\begin{figure}[htbp]
\centering
\includegraphics[width=.8\textwidth]{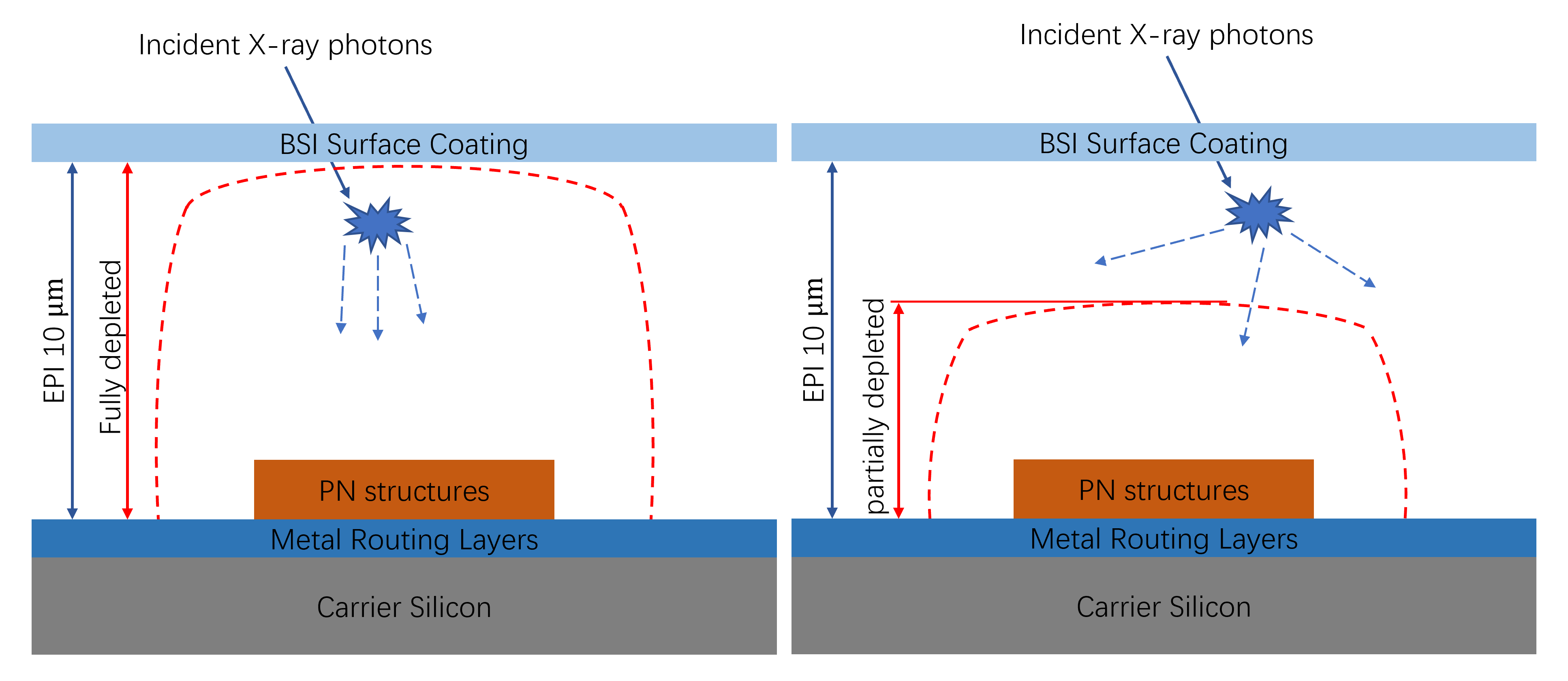}
\caption{\label{fig:pixel_struc} A schematic diagram of the pixel structure of a fully depleted sensor (left) and a partially depleted sensor (right). The charge in partially depleted region has a higher chance to drift to nearby pixels.}
\end{figure}

\begin{figure}[htbp]
\centering
\includegraphics[width=.7\textwidth]{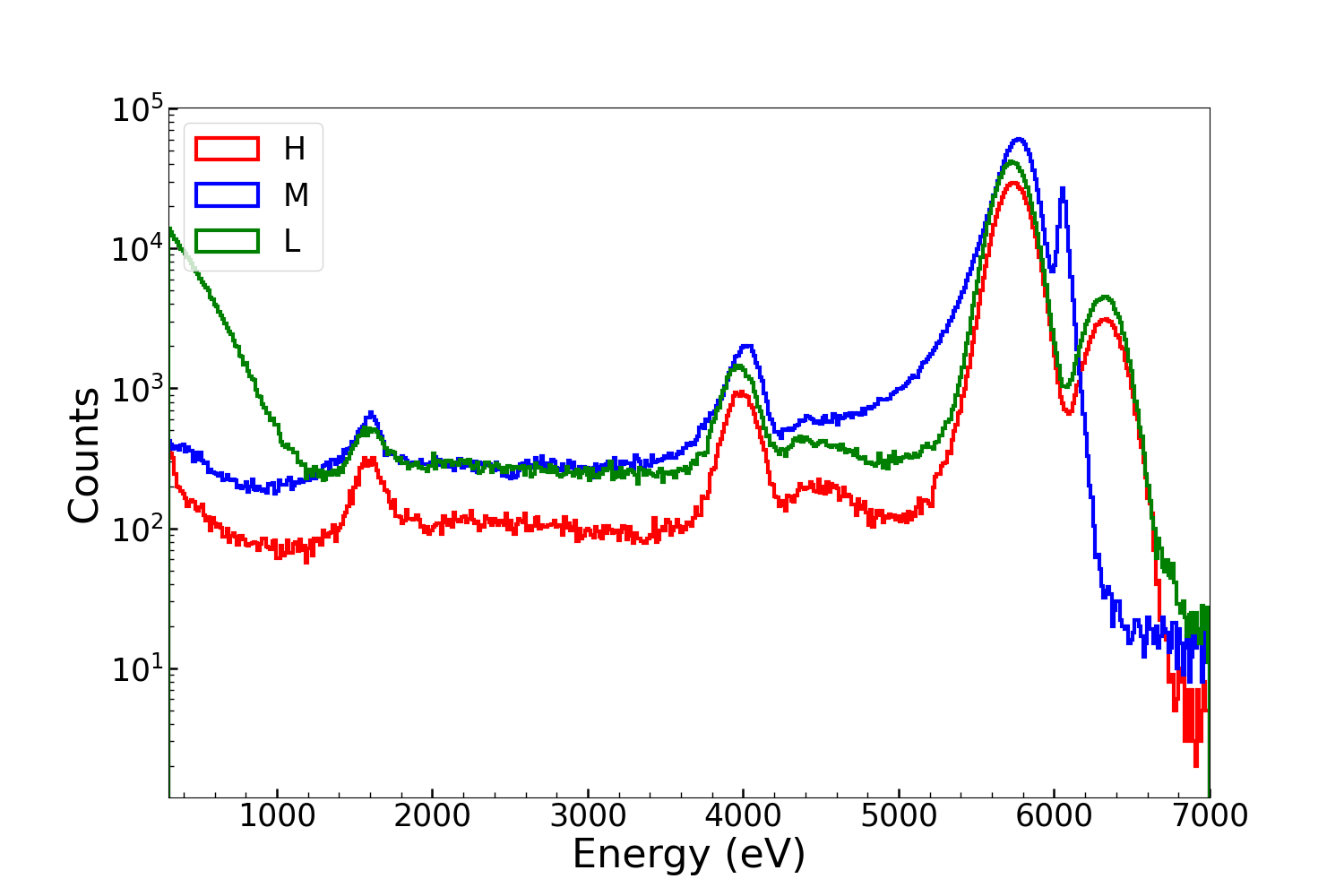}
\caption{\label{fig:spec_1pix} The spectra of single-pixel events obtained by the three sensors. Four peaks (Si $\rm{K_\alpha}$ at 1.74 keV, Si escape peak of Mn $\rm{K_\alpha}$ at 4.16 keV, Mn $\rm{K_\alpha}$ at 5.90 keV, and Mn $\rm{K_\beta}$ at 6.49 keV) can be identified.}
\end{figure}

For the partially depleted sensors M and L, an area of $3\times3$ may be insufficient to include all the electrons caused by an X-ray photon. As shown in the right panel of Figure~\ref{fig:Bar3d}, the electrons can split into a larger area of $7\times7$ in an event detected by sensor L. Figure~\ref{fig:pixel_struc} illustrates the scenario of X-ray photons caught by the sCMOS sensor of different depletion. In a partially depleted sensor, the charge is easier to drift in the neutral region and spread to nearby pixels. This can lead to the charge absorbed by the structure between pixels and explain why serious signal loss could happen in a sensor with lower depletion depth. Among all the events, single-pixel events can give the cleanest spectrum and the best energy resolution. The spectrum of single-pixel events for each sensor is displayed in Figure~\ref{fig:spec_1pix}. The split threshold is set at 50 DN which is above $\rm{10\sigma}$ level. Four peaks can be identified from each spectrum, namely Si $\rm{K_\alpha}$ at 1.74 keV, Si escape peak of Mn $\rm{K_\alpha}$ at 4.16 keV, Mn $\rm{K_\alpha}$ at 5.90 keV, and Mn $\rm{K_\beta}$ at 6.49 keV. The peak location and full width at half maximum (FWHM) which indicates the energy resolution are derived with Gaussian fittings. For single-pixel events, the energy resolution is 204.6 eV, 225.5 eV, and 204.7 eV at 5.90 keV for the sensor H, M and L respectively. Type M sensor shows an unusual noise level along with the X-ray spectrum, which is thought to arise from the individual imperfect structure of the sensor M itself. However, it should be noted that the results are tentative for the H and L sensors. Comparing only the energy resolution of the type H and type L sensors, we can see that the decrease in depletion depth makes little difference in the energy resolution of spectra for the single-pixel events. This indicates that the X-ray photo-electrons can be mostly collected in single-pixel events.

\begin{figure}[htbp]
\centering
\includegraphics[width=.7\textwidth]{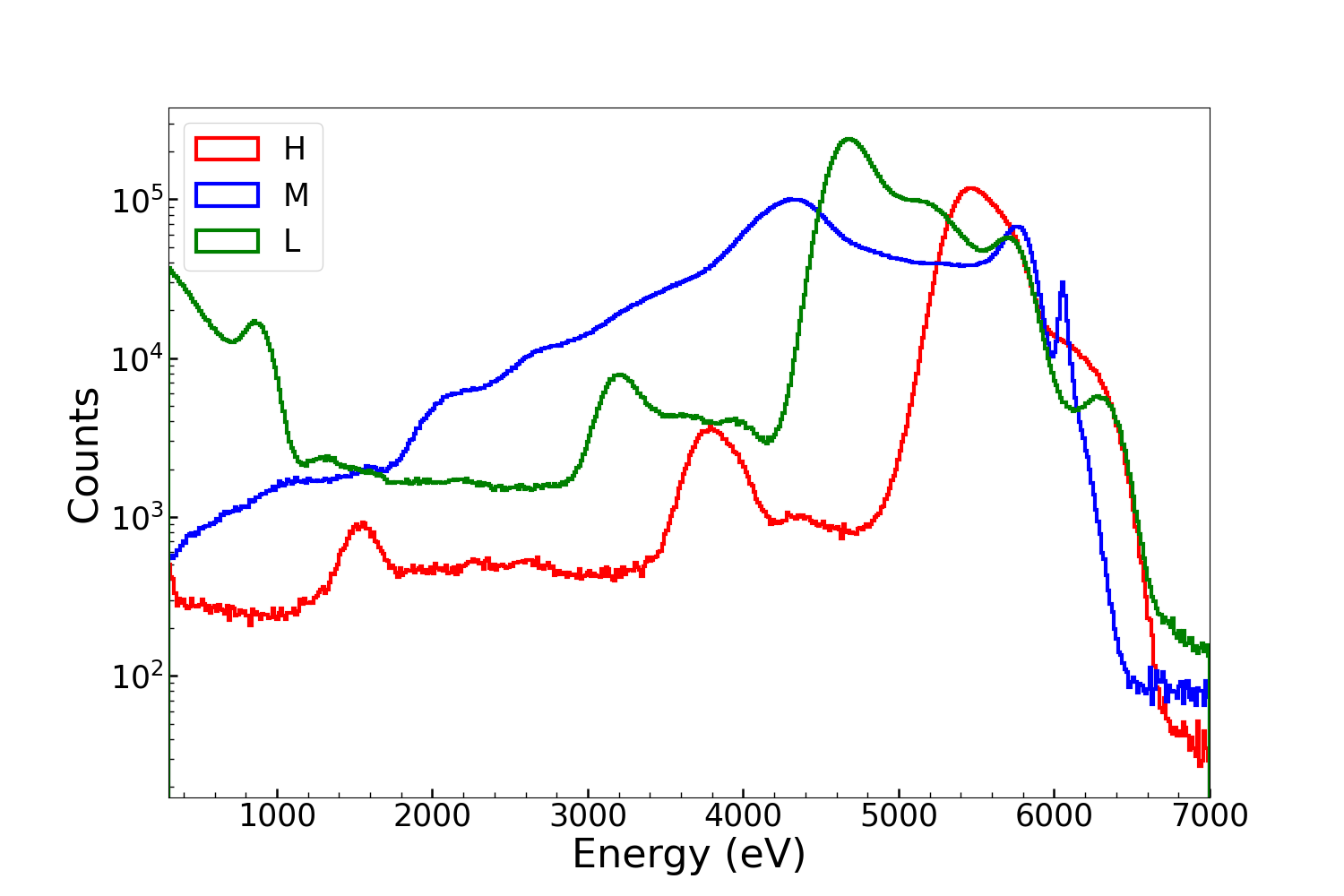}
\caption{\label{fig:spec_7by7} The spectrum of all types of events.}
\end{figure}

Figure~\ref{fig:spec_7by7} shows the spectrum including all types of events for the three sensors. All the charge in a $7\times7$ region of each event is calculated. Compared to the spectrum of single-pixel events in each sensor, the spectra including all the events show distortion and shift of the Mn $\rm{K_\alpha}$ peak, and a shoulder appears near the peak. These indicate that multi-pixel events suffer from signal loss. The shift of the peak location is evidently higher in a sensor with lower depletion depth. This indicates that a lower depletion depth can result in a wider charge dispersion. The rise at low energies (below 1 keV) of the spectrum of the type L sensor is assumed to be the noise peak. More details about the depletion depth and charge dispersion will be discussed in Section~\ref{sec:CG}.

The fraction of different grades of events for the three sensors with split threshold set at 50 DN are listed in Table~\ref{tab:evt_frac}. More multiple-pixel events are found in a sensor with lower depletion depth, which supports the idea that higher depletion depth should result in more single-pixel events. This also suggests that the decrease in depletion depth can lead to more severe charge dispersion.

\begin{figure}[htbp]
\centering
\includegraphics[width=.7\textwidth]{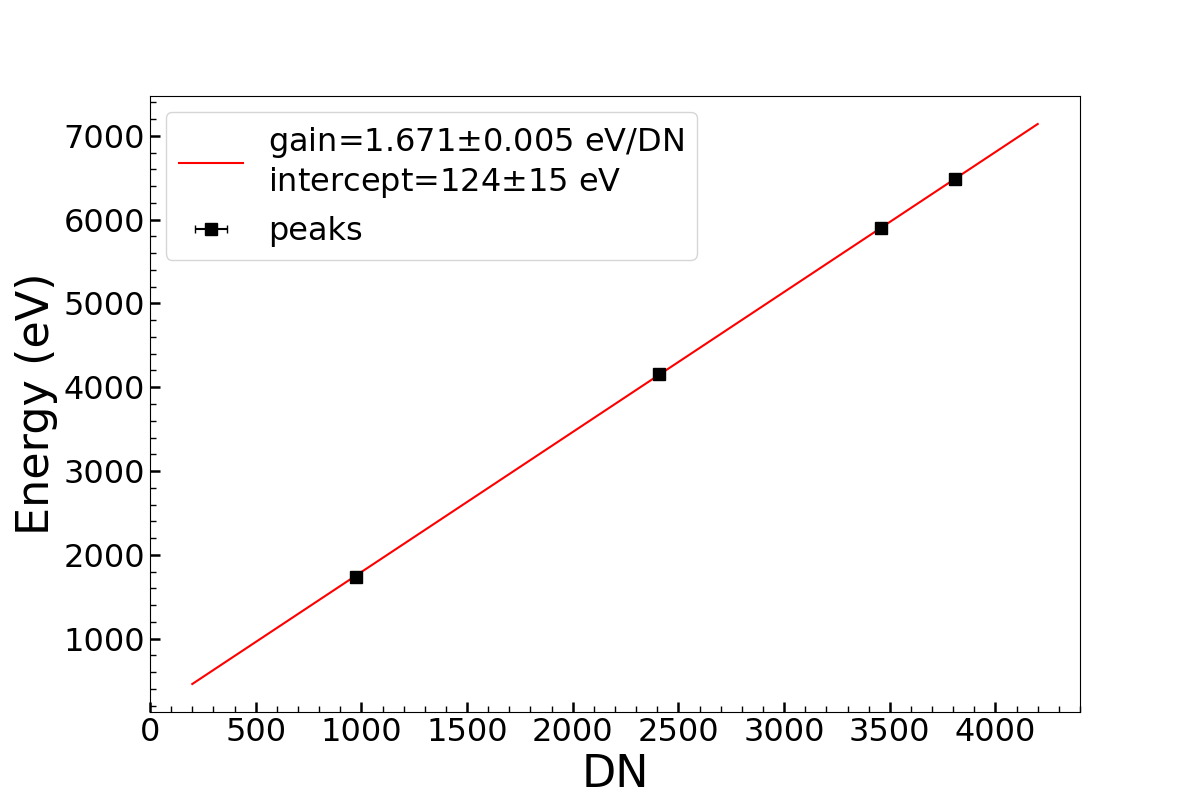}
\caption{\label{fig:gain} A linear fit to derive the conversion gain at the temperature of -20$^\circ \rm{C}$, taking the sensor H as an example.}
\end{figure}

Using the above four peaks in Figure~\ref{fig:spec_1pix}, we performed a linear fit to obtain the conversion gain from DN to eV. Figure~\ref{fig:gain} gives an example of the fits based on the single-pixel events in sensor H. The gain is $1.671\pm0.005$ eV/DN at $-20^\circ \rm{C}$. The gains for the other two sensors were also derived in this way and the values turn out to be $1.502\pm0.016$ eV/DN for the sensor M and $1.657\pm0.006$ eV/DN for the sensor L. These gain values are consistent with those given in the official data sheet.

\begin{table}[htbp]
\centering
\caption{\label{tab:evt_frac} The fraction of different events under fixed split threshold 50 DN at -20$^\circ \rm{C}$.}
\smallskip
\begin{tabular}{c c c c c c}
\hline
Event Type & 1-pixel event & 2-pixel event & 3-pixel event & 4-pixel event & others\\
\hline
H & 15.78$\%$ & 13.08$\%$ & 7.42$\%$ & 30.14$\%$ & 33.57$\%$\\
M & 13.82$\%$ & 8.14$\%$ & 3.51$\%$ & 14.64$\%$ & 59.89$\%$\\
L & 11.03$\%$ & 4.63$\%$ & 1.45$\%$ & 6.95$\%$ & 75.94$\%$\\
\hline
\end{tabular}
\end{table}

\subsection{Center-of-gravity model and depletion depth estimation}
\label{sec:CG}
As shown in Figure~\ref{fig:Bar3d}, a lower depletion depth results in stronger charge dispersion, which means that there would be a discrepancy in the charge distribution of an X-ray event. Thus we introduced the distribution of the center-of-gravity (CG)~\cite{a} of the events detected by the three different sensors, in an attempt to estimate the depletion depth. Only single-pixel events are used considering their relatively complete collection of the photon-electrons. 

We assumed that the signals are well retained within $p_{\rm h}-3\sigma_{\rm h}<s<p_{\rm h}+3\sigma_{\rm h}$, where s represents the DN value of an event, $p_{\rm h}$ is the peak location of the highest Mn $\rm{K_\alpha}$ line (5.90 keV) given by Gaussian fitting and $\sigma_{\rm h}$ is the corresponding $1\rm{\sigma}$ value. The CG of an event is calculated by product of the charge proportion collected in a pixel and the location of the pixel. For example, if an event in a $3\times3$ window like the one shown in the upper right of Figure~\ref{fig:grade} is recorded, the CG of this event will be calculated as $5/3100\times(-1,1)+10/3100 \times(0,1)+...+5/3100\times(1,-1)$.

\begin{figure}[htbp]
\centering 
\includegraphics[width=.32\textwidth]{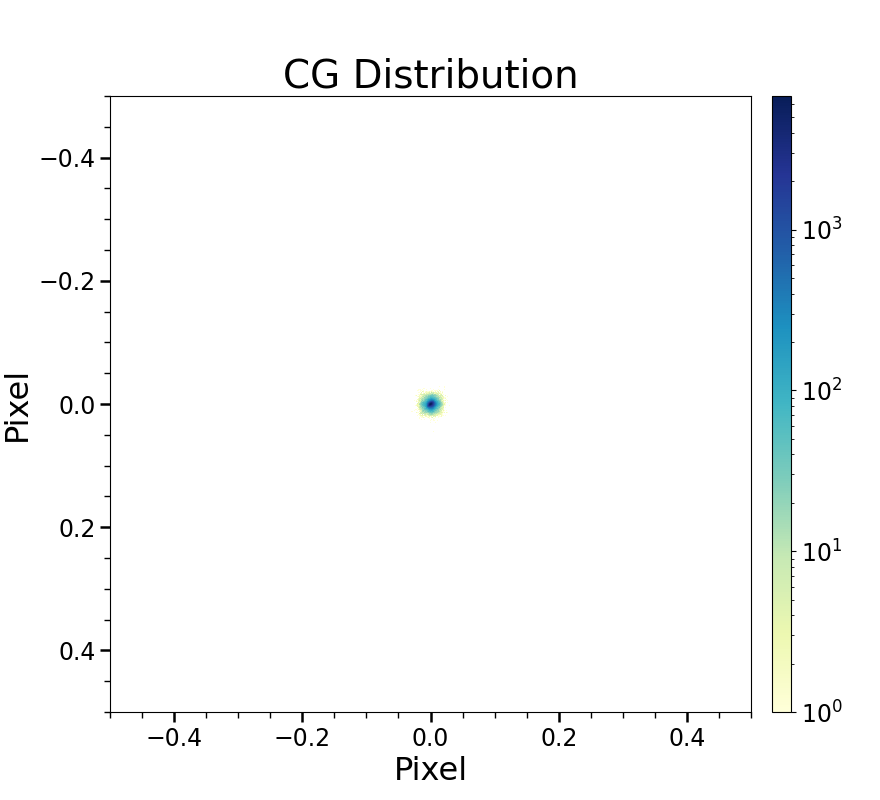}
\includegraphics[width=.32\textwidth]{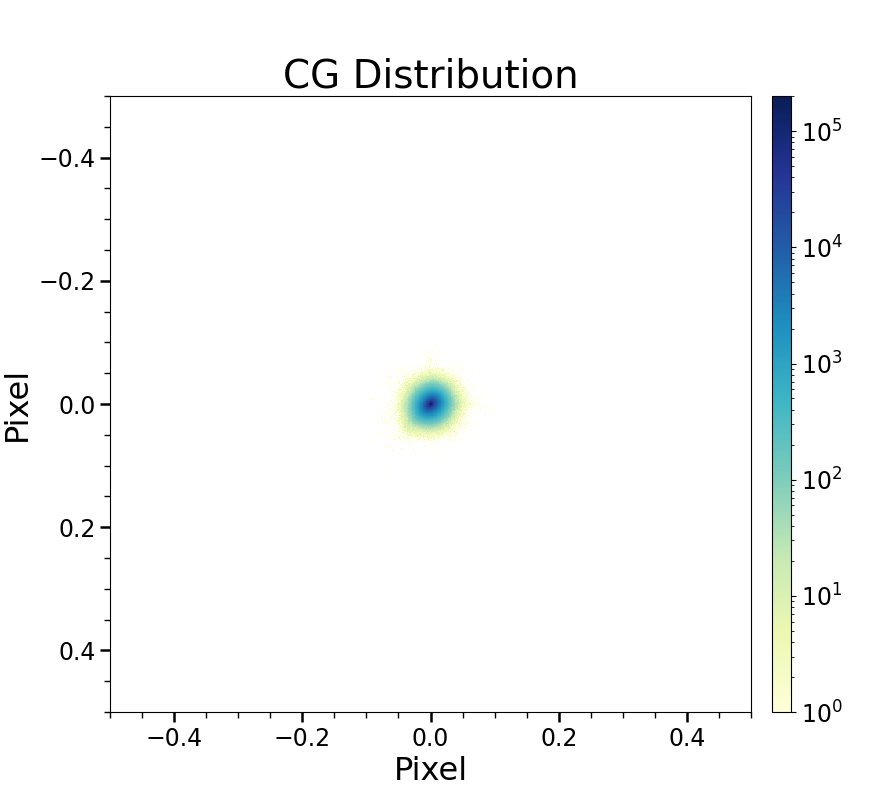}
\includegraphics[width=.32\textwidth]{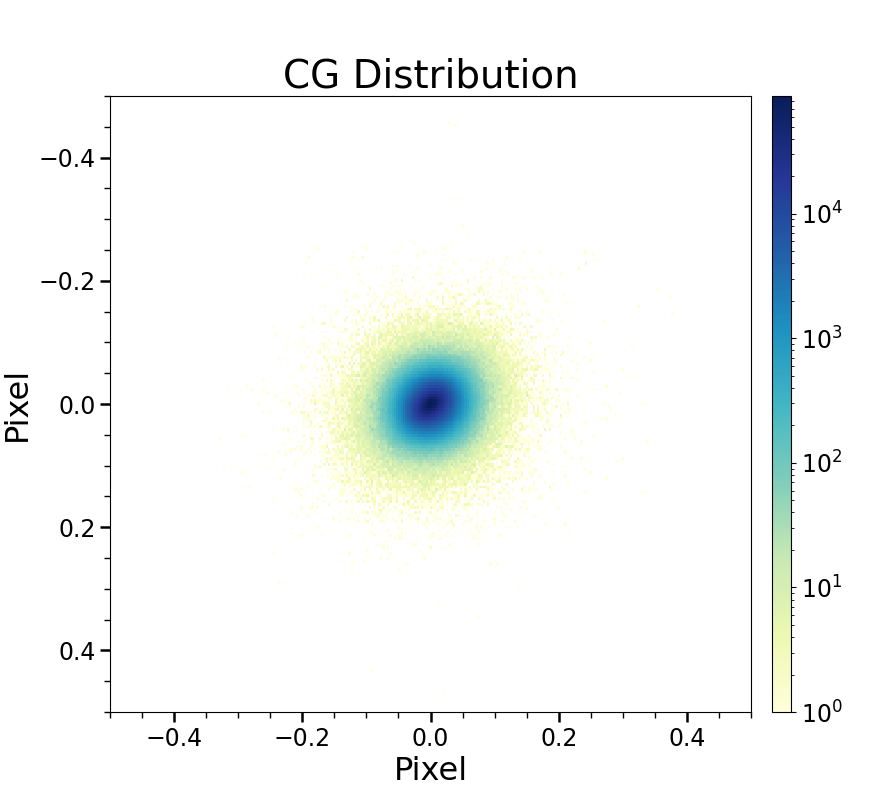}
\caption{\label{fig:CG} The center-of-gravity (CG) distribution map of the three sensors (H, M and L from left to right, respectively). It shows that a lower depletion depth suffers from more severe charge dispersion.}
\end{figure}

For the signal-preserved events described above, we obtained the distribution maps of the CG for the sensor H, M and L respectively. The results are shown in Figure\ref{fig:CG}. It is clear that the CG spreads to a wider range as the depletion depth of the sensor decreases. To further quantify this diffusion, we performed a 2-dimensional Gaussian fitting to each of the distribution, and use the equivalent length of its semi-major axis $\sigma$ to describe the depletion depth. This yields a $\sigma$ value of 0.0025, 0.0032 and 0.0131 for type H, M and L sensors, respectively. We then simply assume that the depletion depth D of a sensor is inversely proportional to its $\sigma$ value. Using this relation and combined with the depletion depth of the fully depleted type H sensor, which is 10.0 $\rm{\mu m}$ according to the fab data sheet, we estimated the depletion depth of the two partially depleted sensors. The resultant depletion depth is 7.8 $\rm{\mu m}$ in type M sensor and 1.9 $\rm{\mu m}$ in type L sensor. It should be noted that this is just a very simple model, and further study has to be carried out by using more of other sCMOS sensors.

\section{Conclusions}
\label{sec:conclusions}
Despite that the sCMOS devices have gained increasing attention and application in X-ray detection, there still remain some challenges such as their limited depletion depth. In this work we tested a small-pixel-sized BSI sCMOS sensor GSENSE2020BSI to examine its basic properties and X-ray performance. It turns out that G2020 works well as an X-ray detector. The fully depleted G2020 sensor reaches a dark current of 0.5 $\rm{e^-/pixel/s}$ and a readout noise of 1.6 e$^-$ at -20$^\circ \rm{C}$. Single-pixel events give the cleanest spectrum and the best energy resolution of 204.6 eV (3.5$\%$) at 5.90 keV.

To further study the effect of depletion depth on the performance, we produced and examined three versions of G2020 with different depletion depths. The results show that the thickness of depletion region makes little difference to the readout noise, dark current level and even the energy resolution for single-pixel events of a sensor. However, sensors with a lower depletion depth suffer from worse charge dispersion and severe charge loss, which is prominent in the spectra of higher or all grades events. An X-ray event of a partially depleted sensor can spread into a $5\times5$, or even $7\times7$ pixel region, which differs from the usual case of $3\times3$ region. We introduced the CG distribution model to study the charge spread and find the distribution to be related to the depletion. Using a 2D Gaussian fit and assuming an inverse proportional relation between the depletion depth and the charge dispersion, we could estimate the depletion depth of the partially depleted sensors with the simple model. More detailed work will be conducted with our other sCMOS sensors. 

In general, both fully depleted and partially depleted sCMOS sensors can be applied for X-ray detection. However, partially depleted type sensors tend to deliver X-ray spectra of less quality, especially for multi-pixel events.  A fully or highly depleted type of sCMOS sensor is thus preferred for application of X-ray spectroscopy.


\acknowledgments
This work is supported by the National Natural Science Foundation of China (grant no. 12173055) and the Chinese Academy of Sciences (grant no. XDA15310100, XDA15310300, XDA15052100).


\end{document}